\documentclass[twocolumn,aps,prb,showpacs]{revtex4}
\usepackage{amsfonts}
\usepackage{amsmath}
\usepackage{graphicx}
\usepackage{bm}
\usepackage{amssymb}
\usepackage{times}
\usepackage{dcolumn}

\makeatletter

\begin{document}

\title{Charged States and Band-Gap Narrowing in Codoped ZnO Nanowires for
Enhanced Photoelectrochemical Responses}
\author{Zhuo Xu}
\author{Qing-Rong Zheng}
\author{Gang Su}
\email[Corresponding author. ]{Email: gsu@gucas.ac.cn}
\affiliation{{Theoretical Condensed Matter Physics and Computational
Materials Physics Laboratory, College of Physical Sciences, Graduate
University of Chinese Academy of Sciences, P.O. Box 4588, Beijing
100049, China}}

\begin{abstract}
By means of first-principles calculations within the density
functional theory, we study the structural and optical properties of
codoped ZnO nanowires and compare them with those of the bulk and
film. It is disclosed that the low negative charged ground states of
nitrogen related defects play a key role in the optical absorption
spectrum tail that narrows the band-gap and enhances the
photoelectrochemical response significantly. A strategy of
uncompensated N, P and Ga codoping in ZnO nanowires is proposed to
produce a red-shift of the optical absorption spectra further than
the exclusive N doping and to get a proper formation energy with a
high defect concentration and a suppressed recombination rate. In
this way, the absorption of the visible light can be improved and
the photocurrent can be raised. These observations are consistent
with the existing experiments, which will be helpful to improve the
photoelectrochemical responses for the wide-band-gap semiconductors
especially in water splitting applications.
\end{abstract}

\maketitle

\section{I\lowercase{ntroduction}}

ZnO is a semiconductor with wide band-gap, and is stable in aqueous
solution. One of the most remarkable applications of ZnO in
photoelectrochemical (PEC) field is that, it can be used to split
water by solar light to produce hydrogen as clean fuel. Besides, ZnO
is also a promising candidate of photoanode material in PEC cells
\cite{book}. For most of wide-gap semiconductors such as ZnO and
TiO$_{2}$, the band-gap is far beyond the threshold of 1.23 eV for
the water splitting, even though the overpotential looses require a
larger threshold energy about 1.7 eV \cite{book-TiO2}. On the other
hand, the wide band-gap semiconductors can merely absorb the
ultraviolet portion of the solar spectra. To enhance the efficiency
of utilizing the solar energy, a key issue is how to narrow the
band-gap of semiconductors below 2.0 eV \cite{E-TiO2} so that the
majority of the visible light can be absorbed. To solve this issue,
a suitable band edge positioning with respect to the potentials of
reduction and oxidation reactions, together with the suppression of
charge recombination centers, should be considered \cite{book}.

There has been a number of studies concerning the above
considerations. Among them, the donor-acceptor codoping in ZnO and
TiO$_{2}$ materials is remarkable. Recent theoretical calculations
show that, it is possible to narrow the band-gap of ZnO bulk by
codoping (Ga+N) or (Al+N) \cite{ZnO3N+Ga}; the band-gaps of
TiO$_{2}$ bulk can be narrowed to be 2.26 eV, 1.6 eV and 0.9 eV for
charge compensated codoping (Mo+C) \cite{TiO23Mo+C}, noncompensated
codoping (Cr+N) and (V+C) \cite{TiO23Cr+N}, respectively.
Experimentally, doping Cu \cite{ZnO2Cu} and codoping (Ga+Cu)
\cite{ZnO2Ga+Cu} in p-type, codoping (Ga+N) \cite{ZnO2Ga+N} or
(Al+N) \cite{ZnO2Al+N} in n-type ZnO films have been realized. In
their optical absorption spectra, there is a long tail from the
absorption peak in the ultraviolet range that persists in the
central visible range, thereby narrowing the band-gap. The key
factors to enhance the PEC responses by codoping Ga and N in ZnO
films are believed to include either producing an excellent
crystallinity, or to narrow the band-gap, or to generate a wide
depletion layer to suppress the recombination rate \cite{ZnO2Ga+N}.
It is also observed that N doped n-type ZnO nanowires (ZnONWs)
\cite{E-ZnO1N} can enhance PEC responses that are similar to those
in codoped ZnO films. A theoretical study on the O-vacancy in ZnONWs
suggests that the neutral and charged states of defects have crucial
contributions to optical absorption in the visible range
\cite{ZnO1Ov}. The charged states of O-vacancy in TiO$_{2}$ bulks
have also been emphasized \cite{TiO23Ov1,TiO23Ov2}.

In this paper, we shall explore the structural and optical
properties of uncompensated N, P and Ga codoped ZnONWs by means of
the first-principles calculations within the density functional
theory (DFT) \cite{DFT}. We have found that the low negative charged
states of N-related defects play a key role in producing the long
tail of the optical absorption spectrum, which narrow the band-gap
and enhance the PEC response significantly. It is also observed that
the uncompensated N, P and Ga codoping can generate a red-shift of
the optical absorption spectra larger than the exclusive N doping,
and a proper formation energy with a high defect concentration and a
suppressed recombination rate can be acquired. These observations
will be useful for the application of ZnONWs in the PEC field (e.g.
the water splitting).

This paper is organized as follows. In Sec. II, we shall briefly
introduce the calculational approach, including the generalized
gradient approximation plus U (GGA+U) correction, and the
extrapolation method to correct the band-gaps. In Sec. III, the
negative charged states and optical absorption spectra of N-doped
ZnONWs will be explored. In Sec. IV, the defect bands and
uncompensated N, P and Ga codoping in ZnONWs to enhance the PEC
response will be discussed. An extrapolation method is applied to
discuss the relevant experimental observations. Finally, a summary
is given.

\section{C\lowercase{alculational} A\lowercase{pproach}}

The calculations are performed using the DFT based VASP code
\cite{VASP}, which employs the pseudopotentials of the projector
augmented wave (PAW) \cite{PAW}, with the GGA in the
Perdew-Burke-Ernzerhof (PBE) form \cite{PBE} for exchange
correlations. The cutoff energy for the plane-wave basis set is 400
eV. The convergence criteria for DFT calculations are 10$^{-4}$ eV
of total energy and 10$^{-3}$ eV/{\AA } of force. All structural
relaxations and formation energies are calculated with spin
polarization, with the Morkhorst-Pack k-points meshes of $4\times
4\times 4$, $5\times 5\times 1$ and $1\times 1\times 7$ for ZnO
bulk, ZnO film and ZnONW, respectively. After a full relaxation, the
density of states (DOS) and the optical absorption spectra of ZnONW
are calculated without spin polarization, and the k-points mesh is
raised to $1\times 1\times 21$.

To obtain the absorption spectra, the frequency dependent dielectric
matrices are calculated by VASP after structural relaxation. The
real and imaginary parts of the dielectric tensor are obtained by
the usual Kramers-Kronig transformation and a summation over empty
states, respectively \cite{optical}, from which the energy dependent
dielectric function along the axial direction of the nanowires and
thus the imaginary part of refractive index can be obtained. The
optical absorption intensity is proportional to the imaginary part
of refractive index.

It is known that the conventional DFT method always underestimates
the band-gap of semiconductors, and the position of the charged
states of the defects with respect to the valence band edge also
tends to be underestimated. The character of the defect bands is a
mixture of the valence and conduction bands, which can be reflected
by the change of the defect bands from GGA to GGA+U calculations
\cite{extrapolation}. Note that in the following, Figs. 2-6 are
calculated by GGA without U correction, while for Fig. 7, the GGA+U
method \cite{U} implemented in VASP code is employed to calculate
the optical absorption spectra for several key cases. Dudarev's
approach \cite{Dudarev} is adopted, where only the difference of the
Coulomb and exchange parameters $(U-J)$ is meaningful.

Inspired by the extrapolation scheme in previous works about ZnO
\cite{extrapolation,extrapolation0}, based on the results from both
GGA and GGA+U with a small value of $(U-J)$ for d electrons, we
extrapolate the calculated band-gap of the ground state of N doped
ultrathin ZnONW to the band-gap of the experimentally synthesized N
doped ZnONW \cite{E-ZnO1N}, and then estimate the band-gaps and the
optical absorption peaks if other defects are doped in the same
ZnONWs in experiments. The extrapolation equation that is similar to
the one given in Ref. \onlinecite{extrapolation} is as follows:
\begin{align}
&E_{g}(D)=E_{g}^{GGA+U}(D)+ \nonumber \\
&\frac{E_{g}^{GGA+U}(D)-E_{g}^{GGA}(D)}%
{E_{g}^{GGA+U}(N_{O})-E_{g}^{GGA}(N_{O})}%
[E_{g}^{exp}(N_{O})-E_{g}^{GGA+U}(N_{O})],
\end{align}
where $N_{O}$ represents substitutionally doping one N atom instead
of O atom, and $D$ represents other defects. Any band-gap $E_{g}$
corresponds to a peak in the optical absorption spectrum, no matter
how it is obtained either from GGA ($E_{g}^{GGA}$) and GGA+U
($E_{g}^{GGA+U}$), or from the experiment ($E_{g}^{exp}$).

Besides the extrapolation method to correct the band-gap, it is
instructive to note that by incorporating U not only for d electrons
but also for s or p electrons, the effective corrections to the
band-gaps and spectra have already been obtained in several cases
\cite{ZnO1Ov,TiO23Ov2,ZnO3Ov}. To aim at giving an approximate but
intuitive view on the absorption spectra, we shall consider several
situations for a comparison in the following [as will be seen in
Figs. 7(a) and (b)], namely, U is incorporated for the d electrons
of Zn, p electrons of O \cite{ZnO1Ov}, as well as the corresponding
electrons of their substitutional dopants.

In a supercell, the formation
energy $E^{f}$ of a charged defect $D$ with
a net charge $q$ can be calculated through the following equation
\cite{formation1,formation2,TiO23Ov1}
\begin{equation}
E^{f}(D^{q})=E^{tot}(D^{q})-E^{tot}(pure)-\Sigma _{i}\Delta n_{i}
\mu_{i}+q\mu _{e},
\end{equation}%
where $E^{tot}(D^{q})$ and $E^{tot}(pure)$ are the total energy with and
without the defects, respectively, $\Delta n_{i}$ is the
change of atomic number, $\mu _{i}$ is the reservoir chemical potentials,
$i= $Zn, O, N, P, Ga, and $\mu _{e}$ is the Fermi level (FL) with
respect to the valence band edge varying between zero and the
band-gap $E_{g}$.

For charged defects, a jellium background is used. However, we note
that the periodic boundary conditions may induce spurious
electrostatic interactions between charged unit cells, possibly
leaving errors in energy from the first-principles calculations
\cite{PBC}. For an aperiodic three-dimensional condensed matter
system, Makov and Payne evaluated the total energy by \cite{PBC}
\begin{equation}
E^{tot}=E_{0}-\frac{q^{2}\alpha}{2L\epsilon}-\frac{2\pi qQ}{3L^{3}
\varepsilon}+\mathcal{O}(L^{-5}),
\end{equation}%
where $E_{0}$, $q$, $\alpha$, $L$, $\epsilon$, and $Q$ is the
desired electrostatic energy of the point defect in the dilute
limit, the ionic charge of the defect, the Madelung constant of the
lattice, the linear dimension of the supercell, the dielectric
constant, and the quadrupole moment independent of $L$,
respectively. The $L^{-1}$ term is the Madelung energy for the
lattice of point charges in the jellium background, and for quasi
one-dimensional system such as nanowires, $L$ corresponds to the
periodicity of the nanowire \cite{1DPBC1}. The $L^{-3}$ term exists
with a quadrupole moment and is inversely proportional to the
supercell volume \cite{PBC}. If the vacuum separation between the
nanowires is fixed and only the axial length $l$ of supercell
changes, the supercell volume is proportional to $l$. More methods
to correct the formation energy have been demonstrated
\cite{1DPBC1,1DPBC2} for several cases of nanowires.

In the present work, to evaluate the effect of the energy errors on
the structural stability, we model the energy correction term
similar to the form of Madelung energy. The total energy is
expressed as
\begin{equation}
E^{tot}=E_{0}+\frac{C}{l},
\end{equation}%
where $C$ is the parameter and is zero for neutral charged
nanowires, and $l$ is the axial length of the supercell. Then, Eq.
(2) is corrected to be
\begin{equation}
E^{f}(D^{q},l)=E^{f}_{0}(D^{q})+\frac{C}{l},
\end{equation}%
\begin{equation}
E^{f}_{0}(D^{q})=E_{0}(D^{q},l)-E_{0}(pure,l)-\Sigma _{i}\Delta n_{i}\mu
_{i}+q\mu _{e}.
\end{equation}%
$E^{f}_{0}(D^{q})$ is the desired energy of the defect in the dilute
limit. For a given charged defect $D^{q}$ with given $\mu _{i}$ and
$\mu _{e}$, $E^{f}_{0}(D^{q})$ and the parameter $C$ of the defect
can be obtained by calculating $E^{f}$($D^{q}$,$l$) with an axial
length $l$ of different unit cells, and then linearly fitting
\cite{fitting} $E^{f}$($D^{q}$,$l$) to the inverse of $l$. Thus, the
error of the defect formation energy $\Delta E^{f}$($D^{q}$,$l$) can
be evaluated by
\begin{equation}
\Delta E^{f}(D^{q},l)=E^{f}_{0}(D^{q})-E^{f}(D^{q},l).
\end{equation}%

The supercells of ZnO bulk, ZnO film and ZnONW contain 128, 144, 96
atoms in total, respectively. The ZnO film is modeled perpendicular
to the axis along the [11$\bar{2}$0] direction, containing 9
central-symmetric layers, 13.3 {\AA } thick. The central three
layers are fixed to the bulk positions while the top and bottom
three layers are allowed to relax without constraints. The supercell
of ZnONW contains two periodic units along the [0001] direction with
diameter of 9.9 {\AA }, and the axial length of the supercell is
10.84 {\AA }. The vacuum distances between the surfaces of
neighboring ZnO film and ZnONW are kept 9.7 and 15.1 {\AA },
respectively. The size of the supercell of ZnONW is $25.0\times
25.0\times 10.84$ {\AA }. The lengths of these three supercells of
ZnO bulk, ZnO film and ZnONW are marked by $2x\times 2y\times 2z$,
$v\times 2y\times 2z$ and $v\times v\times 2z$, respectively. For
each case, $x$, $y$ and $z$ represent the unit lengths of the axial
directions, and $v$ represents the given length of supercell in the
direction with vacuum. By default, these supercells are used to
calculate the formation energies $E^{f}$($D^{q}$,$l$=2$z$) directly
by Eq. (2).

To evaluate the error of the defect formation energy $\Delta
E^{f}$($D^{q}$,2$z$) by Eq. (7) and the linear fitting method
\cite{fitting}, the total energies of a $x\times y\times z$
supercell of ZnO bulk, a $v\times y\times z$ supercell of ZnO film,
$v\times v\times z$ and $v\times v\times 3z$ supercells of ZnONW are
also calculated. The influence of the energy error $\Delta
E^{f}$($D^{q}$,2$z$) will be analyzed in the context.

\section{N\lowercase{egative} C\lowercase{harged} S\lowercase{tates}
\lowercase{of} N\lowercase{-doped} Z\lowercase{n}O
N\lowercase{anowires}}

\begin{figure}[tb]
\includegraphics[width=1.0\linewidth,clip]{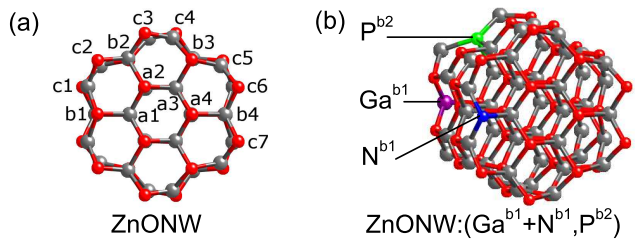}
\caption{(Color online) (a) The structure of the defect free ZnONW
with marks of several positions for doping. (b) The ZnONW of two
periodic units with Ga, N and P dopants at $b1$, $b1$ and $b2$
positions, respectively, represented by
ZnONW:(Ga$^{b1}$+N$^{b1}$,P$^{b2}$). The atoms of Zn, O, N, P and Ga
are colored in gray, red, blue, green and purple, respectively,
throughout the context.} \label{1}
\end{figure}

The structure of the defect free ZnONW is shown in Fig. 1(a), where
several doping positions are marked. To investigate the properties
of the defects in the supercell of ZnONW with two periodic units,
for every position only one of the two units is substitutionally
doped in order to avoid two same dopants occupying the same
position. As an example, ZnONW with Ga, N and P dopants at $b1$,
$b1$ and $b2$ positions, respectively, is shown in Fig. 1(b), marked
as ZnONW:(Ga$^{b1}$+N$^{b1}$,P$^{b2}$), where (Ga$^{b1}$+N$^{b1}$)
represents that a pair of Ga and N dopants at $b1$ are bonded with
each other. Ga substitutes Zn atom, N and P substitute O atoms,
respectively.

\begin{table}[tb]
\caption{The formation energies $E^{f}$ of neutral charged ZnONW:N
under O- and N-rich conditions.}
\begin{tabular}{ccccc}
\hline\hline
position of defect & & & $E^{f}$(eV) \\
\hline
$N^{a1}$ & & & 4.87 \\
$N^{b1}$ & & & 4.72 \\
$N^{c1}$ & & & 4.73 \\
\hline\hline
\end{tabular}
\end{table}

\begin{table}[tb]
\caption{The formation energies $E^{f}$ of neutral charged ZnONW:2N
under O- and N-rich conditions.}
\begin{tabular}{ccccc}
\hline\hline
positions of defects & & & $E^{f}$(eV) \\
\hline
$2N^{b1c1}$ & & & 9.37 \\
$2N^{b1c2}$ & & & 9.39 \\
$2N^{b1b4}$ & & & 9.52 \\
$2N^{b1a2}$ & & & 9.61 \\
$2N^{c2c3}$ & & & 9.53 \\
$2N^{c2c7}$ & & & 9.55 \\
\hline\hline
\end{tabular}
\end{table}

To compare the stability of the structures with different defects,
the formation energies are calculated. For the defect $N_{O}$ in the
ZnONW supercell (ZnONW:N), all $a$, $b$ and $c$ positions are
equivalent, respectively. The formation energies are listed in Table
I. The difference of $E^{f}$ between ZnONW:N$^{b1}$ and
ZnONW:N$^{c1}$ is 0.01 eV, while $E^{f}$ of ZnONW:N$^{b1}$ and
ZnONW:N$^{c1}$ are lower than that of ZnONW:N$^{a1}$ by at least
0.14 eV. For all of these cases the total magnetic moment per
supercell after full relaxation is found to be 1.000 $\mu_{B}$. It
indicates that the surface ($c$) and subsurface ($b$) sites are
energetically favorable for the $N_{O}$ defect. For doping two N
atoms in the supercell (ZnONW:2N), $E^{f}$ of several selected cases
are shown in Table II. Among these cases, ZnONW:2N$^{b1c1}$ is
energetically the most favorable. These results are in general
agreement with the results of a previous work \cite{ZnO1N}. In the
following, only ZnONW:N$^{b1}$ and ZnONW:2N$^{b1c1}$ are discussed
as typical examples with one or two doped N atoms in the ZnONW, if
not specified. For a few interesting cases, more than one
configurations of energetically favorable defects with nearly the
same $E^{f}$ will be considered.

\begin{figure}[t]
\includegraphics[width=1.0\linewidth,clip]{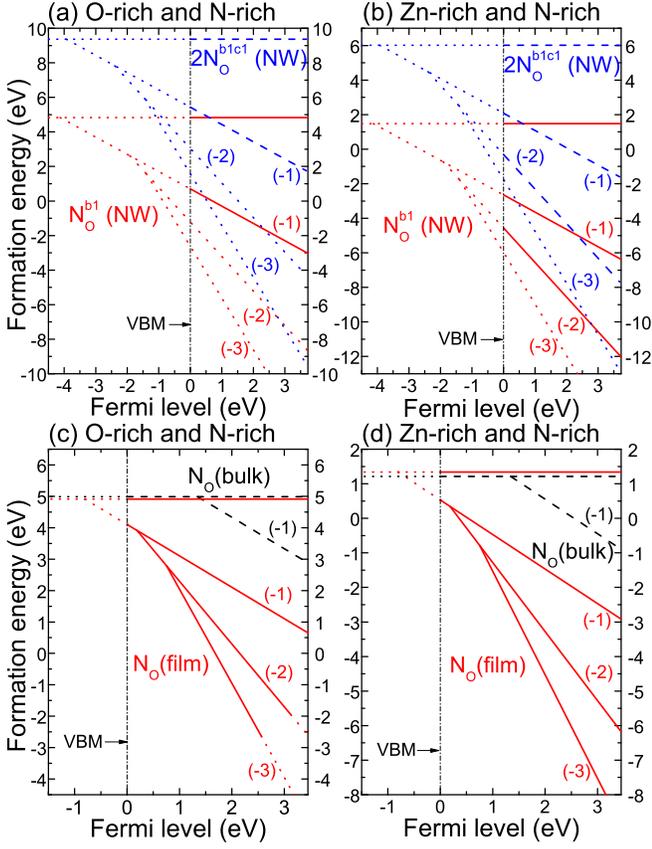}
\caption{(Color online) The formation energy of defect $N_{O}$ in
ZnONW (a) under O- and N-rich, (b) under Zn- and N-rich conditions,
in ZnO bulk and film (c) under O- and N-rich, and (d) under Zn- and
N-rich conditions. The cases of ZnONW:N, ZnONW:2N, ZnO film:N and
ZnO bulk:N are colored in red, blue, red and black, respectively.
Relatively stable states above the VBM are drawn in solid (ZnO
NW/film:N) or dashed (ZnONW:2N/ZnO bulk:N) lines, while the unstable
states above the VBM and all states below the VBM are indicated in
dots, throughout the context.} \label{2}
\end{figure}

Based on Eq. (2), $E^{f}$ of ZnONW:N$^{b1}$ and ZnONW:2N$^{b1c1}$,
and doping one N atom in the supercells of ZnO bulk and film in
different charged states are shown in Figs. 2(a) and (b), (c) and
(d), respectively, where the negative range under the Fermi level
(FL) corresponds to the valence band. As the FL moves from the
valence band maximum (VBM) to the conduction band minimum (CBM), the
semiconductor changes from p-type to n-type. In Fig. 2, as the
band-gap gives the upper limit of the FL, the band-gap of ZnO bulk
is taken as the experimental value of 3.43 eV \cite{U-ZnO3}, while
the band-gaps of ZnO film and ZnONW are approximately set to 3.45
and 3.70 eV, respectively.

For ZnO bulk, the (0/-1) charge transition of $N_{O}$ defect is deep
in the gap, in agreement with the previous observation \cite{p-ZnO}.
On the other hand, the (0/-1) charge transition of ZnO film happens
below the VBM; more than five transitions to negative charged states
of ZnONW appear within the valence band. It is reported both
experimentally and theoretically that the
band of surface O 2p dangling bond of ZnO sits below the VBM. \cite%
{surface1,surface2} This may be true for low dimensional ZnO but not
for all metal-oxide semiconductors \cite{surface3}. So, the charge
transition points below VBM in Fig. 2 correspond to the unoccupied
surface states in the ultrathin ZnO film and ZnONW. As a result, in
contrary to bulk, $N_{O}$ in ZnO film and ZnONW have many negative
charged states in the band-gap, especially under n-type condition
with the FL close to the CBM. These negative charged states are
easier to get negative $E^{f}$ than ZnO bulk.

However, not all of these negative charged states are stable. For ZnONW and
n-type ZnO film, as the defect carries more negative charge, the variation
of $E^{f}$ for adding one more negative charge becomes smaller rapidly. When
the energy difference between different charged states is smaller than the
energy lowering for producing a Zn vacancy ($V_{Zn}$) with the same negative
charge, it will induce spontaneous disintegration from higher negative
states into lower negative states and $V_{Zn}$ defects \cite{DinTiO2}. The
most stable state of $V_{Zn}$ is -2 charged ($V_{Zn}^{2-}$) in spite of
O-rich, Zn-rich, p-type and n-type conditions. For ZnO film,
$E^{f}(V_{Zn}^{2-})$ is -0.76, -7.66, 2.81 and -4.09 eV for O-rich
p-type, O-rich n-type, Zn-rich p-type and Zn-rich n-type extremes,
respectively. For ZnONW, these values are -6.48, -13.86, -3.13, and -10.51
eV, respectively. An approximate criterion to filter the unstable charged
states of $N_{O}$ can be written as
\begin{equation}
E^{f}(N_{O}^{q})+\frac{1}{2}E^{f}(V_{Zn}^{2-})<E^{f}(N_{O}^{q-1})\text{ }
(q\leq 0).
\end{equation}%
If the above inequality is satisfied, the $(q-1)$ charged state will
be unstable; otherwise the charged states are relatively stable
without self-disintegration, as shown in Fig. 2.

For ZnONW:N$^{b1}$ and ZnONW:2N$^{b1c1}$, as manifested in Figs.
2(a) and (b), the relatively stable states free from the spontaneous
disintegration include the neutral and -1 charged states under the
O-rich condition, and 0$\sim $-2 charged states under Zn-rich
condition, among which the ground state in energy is -1 charged
under O-rich condition and -2 charged under Zn-rich condition,
respectively. In general, $E^{f}$ under the O-rich condition is
larger than that under the Zn-rich condition and, it is easier to
get a ground state with a higher negative charge under the Zn-rich
condition than that under the O-rich condition. It is evident that
ZnONW:2N$^{b1c1}$ has a larger $E^{f}$ than ZnONW:N$^{b1}$.

The negative charged states with negative $E^{f}$ play roles as acceptors to
compensate electrons, resulting in high compensation rate and low
photocurrent. The equilibrium concentration of a defect $D$ at temperature
$T$ is given by \cite{formation1,formation2}
\begin{equation}
C(D,q)\varpropto exp[-E^{f}(D,q)/k_{B}T].
\end{equation}%
Obviously, a larger positive $E^{f}$ leads to a lower defect
concentration, and for a negative $E^{f}$, the defect can be viewed
as an active compensation center. So, a small positive $E^{f}$ of
the ground state is preferred, which leads to a high defect
concentration and a low recombination rate. To meet this
requirement, among the ground states of ZnONW:N$^{b1}$ and
ZnONW:2N$^{b1c1}$, $E^{f}$ under the O- and N-rich environments of
ZnONW:(N$^{b1})^{1-}$ at p-type extreme and ZnONW:(2N$^{b1c1})^{1-}$
at n-type extreme are favorable, whose $E^{f}$ are 0.67 and 1.73 eV,
respectively.

The defect energy errors of the charged states are evaluated by Eqs.
(5) and (7) with the linear fitting method. $\Delta
E^{f}$[(N$^{b1}_{O}$)$^{1-}$,2$z$] is found to be -0.43 eV, and the
corresponding parameter $C$ is 4.94 eV$\cdot${\AA }. It means that
after the energy correction, in Fig. 2(a), $E^{f}$ of
ZnONW:(N$^{b1}$)$^{1-}$ under O- and N-rich environments at p-type
extreme drops from 0.67 eV to 0.24 eV. Moreover, $\Delta
E^{f}$[(N$^{b1}_{O}$)$^{2-}$,2$z$] is -1.48 eV, with the parameter
$C$ of 15.50 eV$\cdot${\AA }; $\Delta E^{f}$[(V$_{Zn}$)$^{2-}$,2$z$]
is -1.41 eV, with $C$ of 15.27 eV$\cdot${\AA }. Substituting these
three values of $\Delta E^{f}$ into (8), this inequality is still
satisfied. So, under the O- and N-rich environments, the -1 charged
state of ZnONW:N$^{b1}$ is still the ground state, and at p-type
extreme $E^{f}$ of 0.24 eV will also lead to a high defect
concentration and a low recombination rate. It can be seen that as
the negative charge increases from -1 to -2, $\Delta E^{f}$ of
ZnONW:N$^{b1}$ is enlarged from -0.43 eV to -1.48 eV. For the -3
charged state of ZnONW:N$^{b1}$, $E^{f}$ is expected to be corrected
downward by more than 1.48 eV, and it may turn from the unstable
state to stable state under the Zn- and N-rich environments in Fig.
2(b). However, the -3 charged state is less important and has no
influence on the water splitting. Similarly, after an energy
correction, although $E^{f}$ of ZnONW:(2N$^{b1c1}$)$^{1-}$ is
lowered by a fraction of one electron volt,
ZnONW:(2N$^{b1c1}$)$^{1-}$ under O- and N-rich environments at
n-type extreme is still the only state of ZnONW:2N$^{b1c1}$
favorable for the water splitting.

As a comparison, the charged states of ultrathin ZnO film are also
discriminated by the same criterion in inequality (8), as shown in
Figs. 2(c) and (d). Except for the narrow range near the n-type
extreme under the O-rich environment, the ground state of $N_{O}$
defect in ZnO film is highly negative charged with a negative
$E^{f}$, which is easy to lead to a high recombination rate. If
$E^{f}$ can be lifted from negative to positive, it would suppress
the recombination rate and raise the photocurrent. Normally, the
passivated codoping could lead to a $E^{f}$ larger than the single
doping, as evidenced by the experimental observation that (Ga+N)
codoped ZnO films have a significantly enhanced photocurrent and a
suppressed recombination rate than the solely N doped ZnO film
\cite{ZnO2Ga+N}. The present analysis is effective to explain the
experimental phenomena for both ZnONW and ZnO film. For the -1
charged state, $\Delta E^{f}$ from Eq. (7) is -0.50 and 0.21 eV for
ZnO film and bulk, respectively. However, such energy errors will
not alter the conclusions for ZnO film and bulk.

\begin{figure}[t!]
\includegraphics[width=1.0\linewidth,clip]{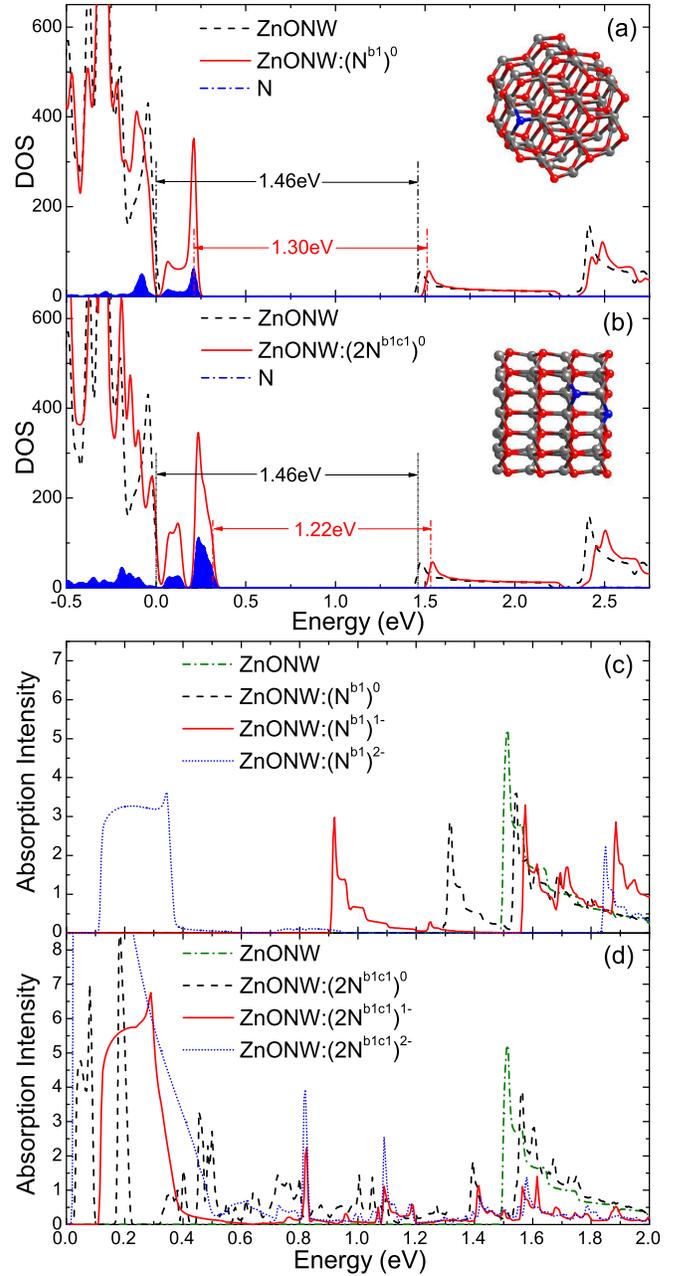}
\caption{(Color online) The DOS of the neutral state of $N_{O}$
doped ZnONW (red solid line): (a) ZnONW:N$^{b1}$ and (b)
ZnONW:2N$^{b1c1}$, in comparison to the pure ZnONW (black dash
line). The partial DOS of N (blue filled area) dopants are also
included. The zero point of energy is set to the VBM of pure ZnONW.
The Fermi level with N dopants are indicated in red short dash line.
The band-gap values are marked. The optical absorption spectra of
$N_{O}$: (c) ZnONW:N$^{b1}$ and (d) ZnONW:2N$^{b1c1}$. The spectrum
of pure ZnONW is indicated by the olive dash-dot line, the spectra
of 0, -1 and -2 charged states are plotted with black dashed line,
red solid line and blue dots, respectively.} \label{1}
\end{figure}

Fig. 3 shows the DOS and optical absorption spectra of
ZnONW:N$^{b1}$ and ZnONW:2N$^{b1c1}$. For both cases, the partial
DOS (PDOS) of N in Figs. 3(a) and (b) reveals that the valence
electrons of N dopants accumulate at the top of the valence band,
resulting in a group of defect bands above the VBM of pure ZnONW.
Compared with the pure ZnONW, the band-gaps of ZnONW:N$^{b1}$ and
ZnONW:2N$^{b1c1}$ are narrowed by 0.16 and 0.24 eV, respectively.

Figs. 3(c) and (d) present the optical absorption spectra of
ZnONW:N$^{b1}$ and ZnONW:2N$^{b1c1}$ with different charged states,
respectively, in comparison with the pure ZnONW. It is noticeable
that even for the negative charged states in Figs. 3(c) and (d), the
defect bands in the band-gap are not so close to the CBM as the gap
is smaller than 0.5 eV. In fact, the spectra of the negative charged
states in the range of 0.0-0.5 eV are contributed by the electron
transition from the main valence bands to the defect bands, or
within the defect bands between the VBM and CBM. So are the spectra
of neutral ZnONW:(N$^{b1})^{0}$ from 0.00 to 1.30 eV, and the
spectra of neutral ZnONW:(2N$^{b1c1})^{0}$ from 0.00 to 1.22 eV. For
the application in the PEC water splitting, only the electron
transition from the valence bands into the conduction bands is
meaningful and effective. Thus, such spectra in the range close to
0.0 eV are ineffective, and will not be discussed. Focusing on the
remaining part of the spectra, the spectra of -1 charged
ZnONW:N$^{b1}$, -1 and -2 charged ZnONW:2N$^{b1c1}$ have red-shifts
more remarkable than those of either their neutral charged states or
the pure ZnONW. A common ground of these low negative charged states
is that their total charges are no larger than that of the
corresponding defect-free nanowires.

Moreover, by incorporating the results from Figs. 2(a) and (b), we
find that ZnONW:(N$^{b1})^{1-}$ at p-type extreme and
ZnONW:(2N$^{b1c1})^{1-}$ at n-type extreme under the O- and N-rich
environments are the two most probable states of the ultrathin ZnONW
with $N_{O}$ defects to perform a nice PEC response. These two
states are ground states with proper formation energies on one hand,
and have the optical absorption spectra with significant red-shifts
on the other hand. If the spectra of neutral charged ZnONWs remain
in the range of ultraviolet light, only these negative charged
states are possible to red-shift the spectra into the visible range,
leaving a series of absorption peaks from the ultraviolet to visible
range.

It is intriguing to note that for N doped n-type ZnONW for PEC water
splitting, the optical absorption spectrum was experimentally
observed to leave a long decaying tail from the ultraviolet to
visible range, where the tail absorbance at 500 nm is approximately
the half of the main absorption peak at 400 nm \cite{E-ZnO1N}. In
experiments it was identified that the dominating defect in ZnONW is
$N_{O}$, with the N concentration of 3.7\% \cite{E-ZnO1N}. Comparing
this experiment to our theoretical results of ZnONW:N$^{b1}$ and
ZnONW:2N$^{b1c1}$, we deduce that the experimentally observed
spectrum with a tail stretching into the visible range can be
attributed to the negative charged ground states of $N_{O}$.

\section{U\lowercase{ncompensated} C\lowercase{odoping}
\lowercase{of} (G\lowercase{a}, N, P) \lowercase{in}
Z\lowercase{n}O N\lowercase{anowires}}

\begin{figure}[t!]
\includegraphics[width=1.0\linewidth,clip]{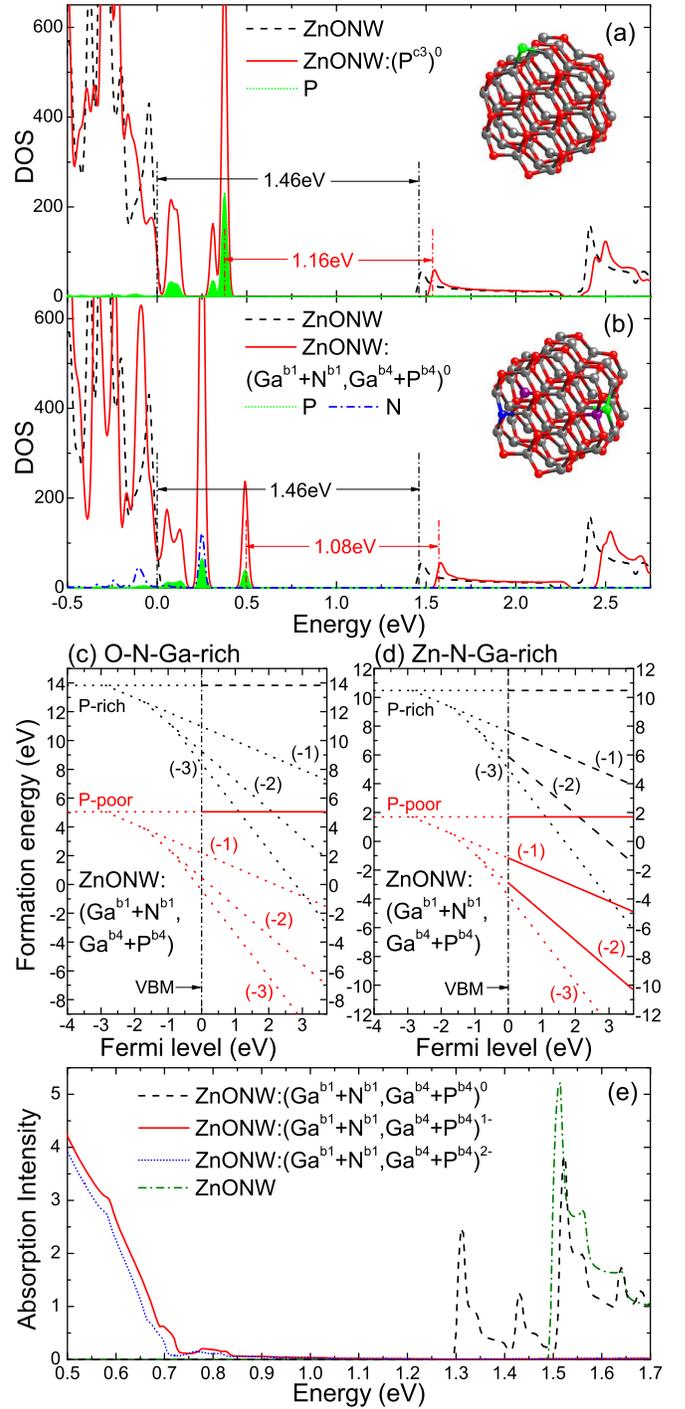}
\caption{(Color online) The DOS of the neutral state for (a)
ZnONW:P$^{c3}$ and (b)
ZnONW:(Ga$^{b1}$+N$^{b1}$,Ga$^{b4}$+P$^{b4}$), including the PDOS of
N (blue) and P (green), which are compared with that of pure ZnONW.
The formation energy of
ZnONW:(Ga$^{b1}$+N$^{b1}$,Ga$^{b4}$+P$^{b4}$) including P-rich
(black) and P-poor (red) cases under (c) O-, N- and Ga-rich, and (d)
Zn-, N- and Ga-rich conditions, respectively. (e) The optical
absorption spectra of pure ZnONW and the charged states of
ZnONW:(Ga$^{b1}$+N$^{b1}$,Ga$^{b4}$+P$^{b4}$).} \label{1}
\end{figure}

To consider the case of $P_{O}$ defect in ZnONW (ZnONW:P), as the
position $c$ is energetically favorable for $P_{O}$, ZnONW:P$^{c3}$
is chosen as an example to investigate the general properties of
ZnONW:P. Fig. 4(a) shows that the defect bands of P situate deeper
than N in the band-gap of pure ZnONW, giving rise to a sharp peak of
DOS, and the band-gap is narrowed from 1.46 eV of the pure ZnONW to
1.16 eV. Aiming at narrowing further the band-gap, it is conceived
that N and P are codoped into ZnONWs. Combining the characteristics
of N and P dopants, the band-gap may be narrowed remarkably by P
while the DOS of the defect bands is maintained to spread from the
VBM as continuously as possible by either N or P dopants.
Accordingly, by incorporating the advantages of passivated codoping
by keeping the crystallinity, controlling the formation energy, and
suppressing the recombination rate \cite{ZnO2Ga+Cu,ZnO2Ga+N}, we try
to codope one (Ga+N) and one (Ga+P) in pair per supercell into
ZnONWs [ZnONW:(Ga+N,Ga+P)].

To examine the above consideration, as an example, the DOS,
formation energy and optical absorption spectrum of
ZnONW:(Ga$^{b1}$+N$^{b1}$,Ga$^{b4}$+P$^{b4}$) are calculated, as
presented in Figs. 4(b)-(e). From Fig. 4(b), one may see that the
band-gap of ZnONW:(Ga$^{b1}$+N$^{b1}$,Ga$^{b4}$+P$^{b4}$) is
narrowed to 1.08 eV, which is smaller than that of ZnONW:P$^{c3}$.
Meanwhile, the gap between the deepest P defect bands and the lower
defect bands of N and P is kept quite narrow.

$E^{f}$ could be manipulated by the chemical potential of P ($\mu
_{P}$) between the values of P-rich and P-poor \cite{PN-ZnO}, where
it is known that the most P-rich potential of $\mu_{P}$ is
calculated from P$_{4}$O$_{10}$ \cite{formation2}, and we choose
P$_{4}$ to obtain a P-poor $\mu_{P}$ approximately. Figs. 4(c) and
(d) show that under the O-rich condition none of the negative
charged states of ZnONW:(Ga$^{b1}$+N$^{b1}$,Ga$^{b4}$+P$^{b4}$) is
stable, and the ground state is neutral charged
[ZnONW:(Ga$^{b1}$+N$^{b1}$,Ga$^{b4}$+P$^{b4}$)$^{0}$] with a
positive $E^{f}$, whereas under the Zn-rich condition the ground
state is -2 charged
[ZnONW:(Ga$^{b1}$+N$^{b1}$,Ga$^{b4}$+P$^{b4}$)$^{2-}$] with a
negative $E^{f}$ at n-type extreme. However, in Fig. 4(e), the
optical absorption spectra of -1 and -2 charged states have no
effective red-shifts from the neutral charged state that can narrow
the gap between the defect bands and the CBM. This is because either
-1 or -2 charged state of
ZnONW:(Ga$^{b1}$+N$^{b1}$,Ga$^{b4}$+P$^{b4}$) has a total charge
larger than that of the defect-free nanowire. Therefore, no matter
whether the formation energy is corrected by Eq. (7) or not,
ZnONW:(Ga+N,Ga+P) is not so useful for the water splitting.

\begin{table}[tb]
\caption{The formation energies $E^{f}$ of neutral charged ZnONW:(Ga,N,P)
under O-, Ga-, P-, and N-rich conditions.}
\begin{tabular}{cccc}
\hline\hline
positions of defects & & $E^{f}(eV)$ \\
\hline
$(Ga^{b1}+N^{b1},P^{a2})$ & & 15.78 \\
$(Ga^{b1}+N^{b1},P^{b2})$ & & 15.75 \\
$(Ga^{b1}+N^{b1},P^{b4})$ & & 15.80 \\
$(Ga^{b1}+N^{b1},P^{b3})$ & & 15.78 \\
$(Ga^{b1}+P^{b1},N^{a2})$ & & 16.2 \\
$(Ga^{b1}+P^{b1},N^{b2})$ & & 16.2 \\
$(Ga^{b1}+P^{b1},N^{b4})$ & & 16.2 \\
$(Ga^{a2}+P^{a2},N^{b1})$ & & 20.1 \\
$(Ga^{b1},N^{b1},P^{b2})$ & & 16.1 \\
$(Ga^{b1}+N^{b1}+P^{a1})$ & & 16.1 \\
$(Ga^{c1}+N^{b1},P^{a2})$ & & 16.3 \\
$(Ga^{a1}+N^{b1}+P^{a2})$ & & 16.2 \\
$(Ga^{a2}+N^{a2},P^{b1})$ & & 16.3 \\
$(Ga^{c3}+N^{c3},P^{b3})$ & & 16.4 \\
\hline\hline
\end{tabular}
\end{table}

To improve the performance, it turns out to take advantage of the
low negative charged ground state with a total charge equal to the
defect-free nanowire, as is in the case of ZnONW:N$^{b1}$. A
strategy can be proposed to promote the PEC responses of ZnONWs by
means of the uncompensated codoping of one (Ga+N) in pair and one
single P dopant per supercell [ZnONW:(Ga+N,P)]. $E^{f}$ for several
ZnONW:(Ga,N,P) cases are listed in Table III. It is seen that the
defects including a pair of (Ga$^{b1}$+N$^{b1}$) and one single P at
$a2$ or $b2$ or $b3$ or $b4$ position are relatively the most
favorable configurations in energy. The difference of $E^{f}$
between these four cases are so small that it is no larger than 0.05
eV. Comparing the cases of (Ga$^{b1}$+N$^{b1}$,P) and
(Ga$^{b1}$+P$^{b1}$,N), one finds that the pair of
(Ga$^{b1}$+N$^{b1}$) is energetically more favorable than
(Ga$^{b1}$+P$^{b1}$). In presence of the (Ga$^{b1}$+N$^{b1}$) pair,
the P dopants at $c$ positions are unstable, because $E^{f}$ in
these cases is usually larger than 200 eV. The three typical
examples such as ZnONW:(Ga$^{b1}$+N$^{b1}$,P$^{b2}$),
ZnONW:(Ga$^{b1}$+N$^{b1}$,P$^{a2}$) and
ZnONW:(Ga$^{b1}$+N$^{b1}$,P$^{b4}$) are energetically favorable,
whose total magnetic moments per supercell after full relaxation are
found to be 1.093, 1.158 and 1.146 $\mu_{B}$, respectively.

\begin{figure}[t!]
\includegraphics[width=1.0\linewidth,clip]{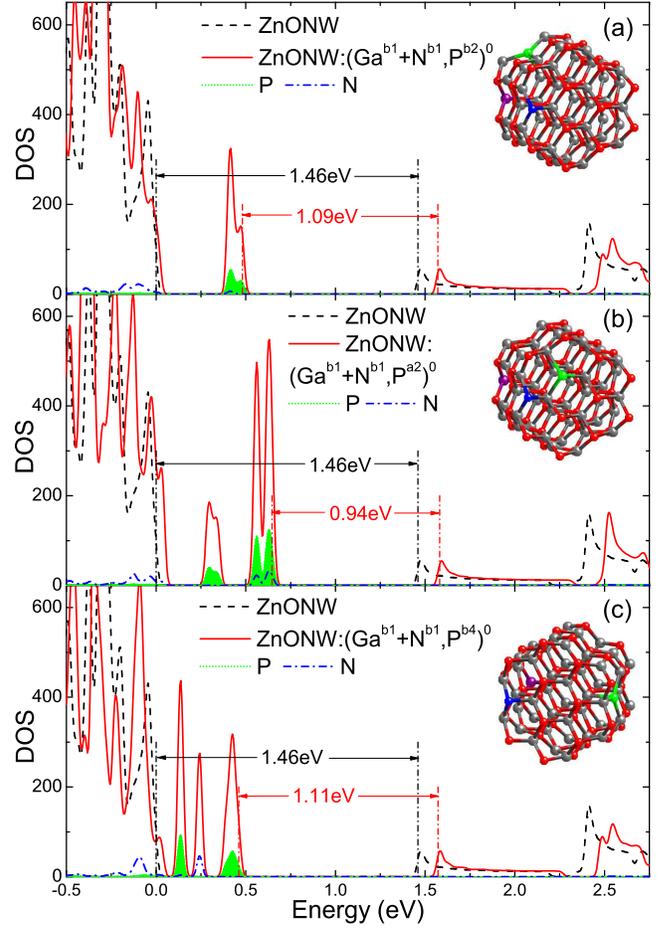}
\caption{(Color online) The DOS of the neutral state for (a)
ZnONW:(Ga$^{b1}$+N$^{b1}$,P$^{b2}$), (b)
ZnONW:(Ga$^{b1}$+N$^{b1}$,P$^{a2}$) and (c)
ZnONW:(Ga$^{b1}$+N$^{b1}$,P$^{b4}$), including the PDOS of N and P,
which are compared with that of pure ZnONW.}
\label{1}
\end{figure}

\begin{figure}[t!]
\includegraphics[width=1.0\linewidth,clip]{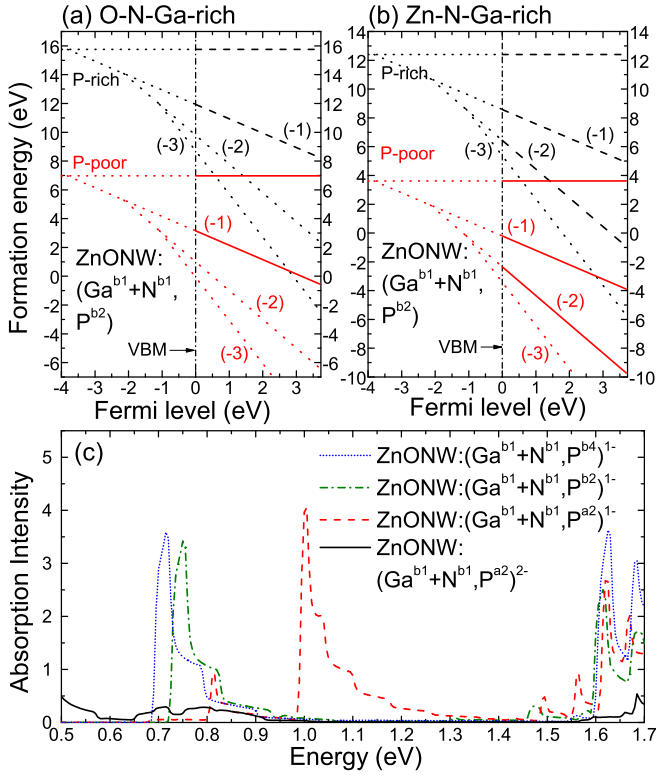}
\caption{(Color online) The $E^{f}$ of
ZnONW:(Ga$^{b1}$+N$^{b1}$,P$^{b2}$) including P-rich (black) and
P-poor (red) cases under (a) O-, N- and Ga-rich, and (b) Zn-, N- and
Ga-rich conditions, respectively. (c) The optical absorption spectra
of ZnONW:(Ga$^{b1}$+N$^{b1}$,P$^{b4}$)$^{1-}$ (blue dots),
ZnONW:(Ga$^{b1}$+N$^{b1}$,P$^{b2}$)$^{1-}$ (olive dash-dot line),
ZnONW:(Ga$^{b1}$+N$^{b1}$,P$^{a2}$)$^{1-}$ (red dash line) and
ZnONW:(Ga$^{b1}$+N$^{b1}$,P$^{a2}$)$^{2-}$ (black solid line).}
\label{1}
\end{figure}

As shown in Figs. 5, all the band-gaps of these three cases
are narrower than those of ZnONW:N$^{b1}$, ZnONW:2N$^{b1c1}$ and the
pure ZnONW, but are close to the band-gap value of
ZnONW:(Ga$^{b1}$+N$^{b1}$,Ga$^{b4}$+P$^{b4}$). For
ZnONW:(Ga$^{b1}$+N$^{b1}$,P$^{a2}$) and
ZnONW:(Ga$^{b1}$+N$^{b1}$,P$^{b4}$), there are also defect bands
spreading between the VBM and the deepest defect peak of DOS.

In Figs. 6(a) and 6(b), the ground states of
ZnONW:(Ga$^{b1}$+N$^{b1}$,P$^{b2}$) under the O-rich and Zn-rich
environments are -1 [ZnONW:(Ga$^{b1}$+N$^{b1}$,P$^{b2}$)$^{1-}$] and
-2 [ZnONW:(Ga$^{b1}$+N$^{b1}$,P$^{b2}$)$^{2-}$] charged,
respectively. $E^{f}$ of ZnONW:(Ga$^{b1}$+N$^{b1}$,P$^{b2}$)$^{1-}$
is -0.54 eV at n-type extreme under the O-, N-, Ga-rich and P-poor
environments [Fig. 6(a)]; $E^{f}$ of
ZnONW:(Ga$^{b1}$+N$^{b1}$,P$^{b2}$)$^{2-}$ is -0.94 eV at n-type
extreme under the Zn-, N-, Ga- and P-rich environments [Fig. 6(b)].
For these two states, evaluating the formation energy error from Eq.
(7), $\Delta E^{f}$[(Ga$^{b1}$+N$^{b1}$,P$^{b2}$)$^{1-}$,2$z$] is
-0.46 eV with the parameter $C$ of 4.99 eV$\cdot${\AA }; and $\Delta
E^{f}$[(Ga$^{b1}$+N$^{b1}$,P$^{b2}$)$^{2-}$,2$z$] is -1.53 eV with
$C$ of 15.88 eV$\cdot${\AA }. Thus, after counting the correction on
energy, $E^{f}$ changes from -0.54 eV to -1.00 eV, and from -0.94 eV
to -2.47 eV. No matter with or without the energy correction, these
two cases are kept to be in the ground states and can easily produce
a small positive $E^{f}$ in the ground state by manipulating the
n-type extreme and the chemical potential $\mu _{P}$. They are
energetically preferable for a high defect concentration and a low
recombination rate. $E^{f}$ of ZnONW:(Ga$^{b1}$+N$^{b1}$,P$^{a2}$)
and ZnONW:(Ga$^{b1}$+N$^{b1}$,P$^{b4}$) are similar to that of
ZnONW:(Ga$^{b1}$+N$^{b1}$,P$^{b2}$) because of a very small $E^{f}$
difference among them.

Fig. 6(c) manifests that for
ZnONW:(Ga$^{b1}$+N$^{b1}$,P$^{b4}$)$^{1-}$,
ZnONW:(Ga$^{b1}$+N$^{b1}$,P$^{b2}$)$^{1-}$ and
ZnONW:(Ga$^{b1}$+N$^{b1}$,P$^{a2}$)$^{1-}$, the leftmost absorption
peak with an intensity larger than 1.0 lies at 0.72, 0.75 and 1.00
eV, respectively. Among them, the spectra of
ZnONW:(Ga$^{b1}$+N$^{b1}$,P$^{b4}$)$^{1-}$ and
ZnONW:(Ga$^{b1}$+N$^{b1}$,P$^{b2}$)$^{1-}$ have further red-shift
from those of ZnONW:(N$^{b1}$)$^{1-}$ and
ZnONW:(2N$^{b1c1}$)$^{1-}$. The peaks around 1.62 eV correspond to
the band-gap from the main VBM to the main CBM, in spite of the
defect bands. In contrast,
ZnONW:(Ga$^{b1}$+N$^{b1}$,P$^{a2}$)$^{2-}$ gives no efficient
band-gap narrowing, so are the -2 charged states of the other two
cases.

\begin{figure}[t]
\includegraphics[width=1.0\linewidth,clip]{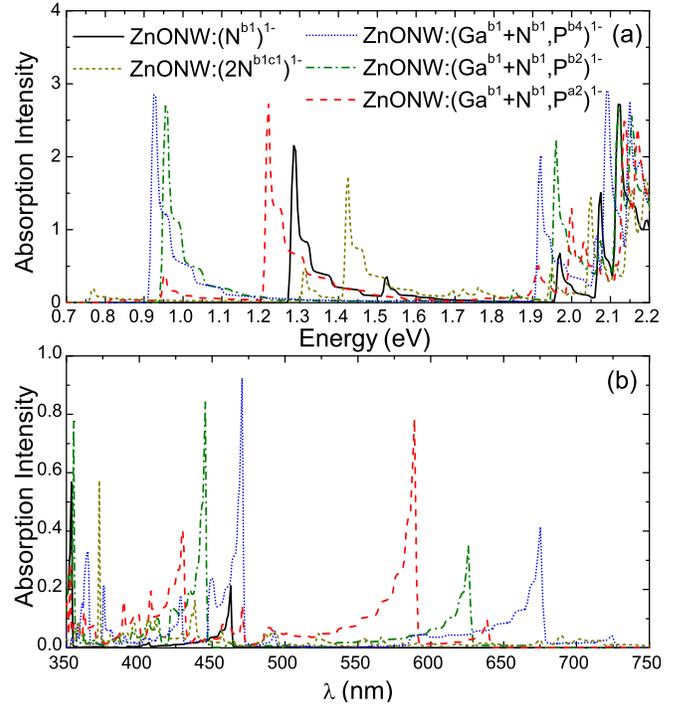}
\caption{(Color online) The optical absorption spectra of
ZnONW:(Ga$^{b1}$+N$^{b1}$,P$^{b4}$)$^{1-}$ (blue dots),
ZnONW:(Ga$^{b1}$+N$^{b1}$,P$^{b2}$)$^{1-}$ (olive dash-dot line),
ZnONW:(Ga$^{b1}$+N$^{b1}$,P$^{a2}$)$^{1-}$ (red dash line),
ZnONW:(N$^{b1}$)$^{1-}$ (black solid line) and
ZnONW:(2N$^{b1c1}$)$^{1-}$ (dark yellow short-dash line), calculated
with GGA+U at (a) $(U-J)$=5.0 eV for d electrons only, and (b)
$(U-J)$=18.0 eV and 15.0 eV for d and p electrons, respectively.}
\label{1}
\end{figure}

Due to the severe underestimation of band-gaps by the DFT, all the
results above calculated within the GGA cannot quantitatively
reflect the realistic situation in PEC responses. For this purpose,
the GGA+U and extrapolation methods that are introduced in Section
II are employed here. When U correction of $(U-J)$=5.0 eV for the
d electrons is added to the GGA calculation,
the band-gaps and spectra are partially corrected,
as shown in Fig. 7(a). For
ZnONW:(Ga$^{b1}$+N$^{b1}$,P$^{b4}$)$^{1-}$,
ZnONW:(Ga$^{b1}$+N$^{b1}$,P$^{b2}$)$^{1-}$,
ZnONW:(Ga$^{b1}$+N$^{b1}$,P$^{a2}$)$^{1-}$, ZnONW:(N$^{b1}$)$^{1-}$
and ZnONW:(2N$^{b1c1}$)$^{1-}$, the leftmost absorption peak with an
intensity larger than 1.0 lies at 0.93, 0.96, 1.22, 1.29 and 1.42
eV, respectively.

From the preceding results, we unclose that under the O-, N-,
Ga-rich, P-poor and n-type environments, the uncompensated (Ga, N,
P) codoping cases such as
ZnONW:(Ga$^{b1}$+N$^{b1}$,P$^{b4}$)$^{1-}$,
ZnONW:(Ga$^{b1}$+N$^{b1}$,P$^{b2}$)$^{1-}$ and
ZnONW:(Ga$^{b1}$+N$^{b1}$,P$^{a2}$)$^{1-}$ are energetically
favorable, whose total charge is equal to that of the defect-free
nanowire. In general, their optical absorption spectra have larger
red-shifts and higher absorption peaks than the cases of exclusive N
doping. Meanwhile, the (Ga, N, P) codoping cases have proper
formation energy that can lead to a high defect concentration and a
low recombination rate. Both factors are much favored in enhancing
the photocurrent in the PEC water splitting.

To show the improvement of the uncompensated codoping of (Ga, N, P)
compared with the exclusive N dopoing in the ZnONWs, in a way
approaching to the experiment, an approximate estimation is
performed by utilizing the extrapolation equation (1). The
calculated results by GGA and GGA+U with a $(U-J)$ of 5.0 eV for d
electrons can be extrapolated to the experimental ones.
ZnONW:(N$^{b1}$)$^{1-}$ is taken as the standard defect of $N_{O}$
in Eq. (1), and $E_{g}^{exp}$ is about 3.11 eV, which corresponds to
the red-shifted peak of the N doped ZnONW at about 400 nm in the
UV-vis spectra in the experiment \cite{E-ZnO1N}. Consequently, the
resulted $E_{g}$ of ZnONW:(Ga$^{b1}$+N$^{b1}$,P$^{b4}$)$^{1-}$,
ZnONW:(Ga$^{b1}$+N$^{b1}$,P$^{b2}$)$^{1-}$ and
ZnONW:(Ga$^{b1}$+N$^{b1}$,P$^{a2}$)$^{1-}$ is 1.96, 1.99 and 2.30
eV, corresponding to 633, 623 and 540 nm, respectively.

Besides the extrapolation method, to give another direct and
intuitive illustration, U correction can be added for both d and p
electrons. When $(U-J)$ is assumed as 18.0 eV for d electrons and
15.0 eV for p electrons, respectively, the long-wavelength end of
the optical absorption spectra of the exclusive N doping is
manipulated to stop just at the range of violet and blue light,
which is close to the experiment \cite{E-ZnO1N}. Such a setting of
GGA+U focuses on directly correcting the band-gap to approach the
experimental value, while the expense is that, the inner band
structures are disturbed more greatly than the extrapolation method
with a small U added only for d electrons. As shown in Fig. 7(b), at
the long-wavelength end of the spectra of ZnONW:(N$^{b1}$)$^{1-}$
and ZnONW:(2N$^{b1c1}$)$^{1-}$, the last peak with an absorption
intensity larger than 0.1 locates at 463 and 438 nm, respectively.
In contrast, the last peak with an absorption intensity larger than
0.3 of ZnONW:(Ga$^{b1}$+N$^{b1}$,P$^{b4}$)$^{1-}$,
ZnONW:(Ga$^{b1}$+N$^{b1}$,P$^{b2}$)$^{1-}$ and
ZnONW:(Ga$^{b1}$+N$^{b1}$,P$^{a2}$)$^{1-}$ locates at 675, 623 and
589 nm, respectively. Generally, the wavelengths of these three
peaks are close to and consistent with those obtained from the
extrapolation method.

Revealed from the estimations either by the extrapolation method or
by directly adding U for both d and p electrons,
the absorption spectra of the exclusive N doping in the ZnONWs
can enter the range of violet light but hard to reach the center of
the visible range, which just manifests the case in the experiment
\cite{E-ZnO1N}. Judging also from the above results by either of
the two estimation methods, one may see that for both N doping and
uncompensated (Ga, N, P) codoping in ZnONWs,
the absorption peaks from about 400 nm to larger than 600 nm
including their tails can cover nearly the whole visible light range,
which may thus lead to the raise of the photocurrent.

\section{S\lowercase{ummary}}

In summary, by means of the first-principles calculations within the
density functional theory, the charged states and band-gap narrowing
in different doped and codoped ZnONWs are investigated and compared
with those of ZnO bulk and film. It is found that the charged states
of nitrogen related defect, particularly the low negative charged
ground state with a total charge no larger than the defect-free
nanowire, play a key role for the PEC response, which leads to a
long tail in the optical absorption spectrum. A design of
uncompensated codoping of N, P and Ga in ZnONWs is proposed, which
could give an evident enhancement of the PEC response. These
observations will in general be useful for improving the PEC
response particularly for the water splitting of wide-band-gap
semiconductors.

\acknowledgments

All calculations are completed on the supercomputer NOVASCALE7000 in
Computer Network Information Center (Supercomputing center) of
Chinese Academy of Sciences and MagicCube (DAWN5000A) in Shanghai
Superconputer Center. This work is supported in part by the NSFC
(Grant Nos. 90922033, 10934008, and 10974253), the MOST of China
(Grant No. 2012CB932901) and the CAS.

\end{document}